\documentclass[aps, preprint, longbibliography]{revtex4-1}
\usepackage{graphicx}
\usepackage{amssymb}
\usepackage{amsmath}

\begin{document}
	
	
	\title{Characterizing vortex tangle properties in steady-state He~II counterflow using particle tracking velocimetry}
	\author{Brian Mastracci}
	\author{Wei Guo}
	\email[Author to whom correspondence should be addressed. Electronic mail: ]{wguo@magnet.fsu.edu}
	\affiliation{National High Magnetic Field Laboratory, 1800 E Paul Dirac Dr., Tallahassee, FL 32310, USA}
	\affiliation{Department of Mechanical Engineering, Florida State University, 2525 Pottsdamer St., Tallahassee, FL 32310, USA}
	
	\date{\today}
	
	\begin{abstract}
	Historically, there is little faith in particle tracking velocimetry (PTV) as a tool to make quantitative measurements of thermal counterflow in He~II, since tracer particle motion is complicated by influences from the normal fluid, superfluid, and quantized vortex lines, or a combination thereof. Recently, we introduced a scheme for differentiating particles trapped on vortices (G1) from particles entrained by the normal fluid (G2). In this paper, we apply this scheme to demonstrate the utility of PTV for quantitative measurements of vortex dynamics in He~II counterflow. We estimate $\ell$, the mean vortex line spacing, using G2 velocity data, and $c_2$, a parameter related to the mean curvature radius of vortices and energy dissipation in quantum turbulence, using G1 velocity data. We find that both estimations show good agreement with existing measurements that were obtained using traditional experimental methods. This is of particular consequence since these parameters likely vary in space, and PTV offers the advantage of spatial resolution. We also show a direct link between power-law tails in transverse particle velocity probability density functions (PDFs) and reconnection of vortex lines on which G1 particles are trapped.
	\end{abstract}
	
	
	\maketitle
	
	\section{Introduction}
	\label{sec:Intro}
	Whole field flow visualization has become a popular research tool for He~II~\cite{Guo2014}, the superfluid phase that occurs in \textsuperscript{4}He at temperatures below about 2.17~K. One common visualization method, particle tracking velocimetry (PTV), tracks the locations of individual micron-sized solidified hydrogen or deuterium tracer particles suspended in the flow field throughout a sequence of photographs. It is trivial to obtain an ensemble of velocity measurements from these time-resolved sequences of particle locations, which can be used, in theory at least, to characterize quantitatively the fluid behavior.
    
    In practice, extraction of reliable, quantitative information from particle velocity measurements has been elusive due to the non-classical mechanics of He~II. The two fluid model of Tisza and Landau describes it as two interpenetrating and fully miscible fluid components~\cite{Tisza1938,Landau1941}. The normal fluid behaves more or less classically, and saturates the He~II system at the phase transition temperature $T_\lambda\approx2.17$~K. It entrains tracer particles by viscous drag. The superfluid component, which saturates the two-fluid system below about 1~K, is inviscid and carries no entropy, but still influences particle motion through inertial and added mass effects~\cite{Sergeev2009}. Furthermore, circulation in the superfluid is confined to quantized vortex lines, each with a single quantum of circulation $\kappa\approx10^{-8}$~m\textsuperscript{2}/s about a core $\xi_0\approx0.1$~nm in diameter. Pressure gradient in the vicinity of each vortex line can attract and trap particles~\cite{Parks1966,Bewley2006}, though once trapped, they have a tendency to slide along the vortex core under the influence of drag exerted by the normal fluid~\cite{Kivotides2008c,Mineda2013}. 
    
    Application of PTV to He~II becomes increasingly complicated in studies of thermal counterflow, a heat transfer mechanism unique to He~II. In response to a thermal stimulus, the normal fluid carries entropy away from the heat source with velocity $v_n$ proportional to the heat flux $q$, while the superfluid moves toward it at $v_s$ such that there is no net mass transfer. As $q$ increases the two fluids can become independently turbulent~\cite{Marakov2015}. Turbulence in the superfluid manifests as a random tangle of quantized vortex lines~\cite{VinenIII}, and a non-classical form of turbulence arises in the normal fluid~\cite{Marakov2015} due to a force of mutual friction that arises from interactions with the vortex tangle~\cite{Hall1956b}. 
    
    Since tracer particles interact with both fluid components, a major challenge when applying PTV to thermal counterflow is determining what influences the motion of an observed particle at a given time, so that the behavior of the underlying flow field can be interpreted correctly~\cite{Guo2013,Vinen2014,Marakov2015,Gao2015}. Until recently analysis was confined to qualitative characterizations: evolution of particle motion in response to applied heat flux~\cite{Chagovets2011}, or of statistical distributions of particle kinematics in response to image acquisition rate~\cite{LaMantia2014a,LaMantia2014b}. A newer visualization technique employing metastable He\textsubscript{2}* eximers as tracer particles avoids this ambiguity issue, since the eximers are not trapped on vortices above about 1~K~\cite{Gao2015}. However, as a compromise, information about the vortex dynamics cannot be obtained, and thus far this method yields information about the flow velocity in one dimension only.
    
    Recently, we studied particle motion in thermal counterflow across a wide heat flux range using PTV, and found that, indeed, particles moving under the influence of relatively high heat flux, to which we give the name G3, are constantly affected by both the normal fluid and vortex lines. However, for relatively low heat flux, we devised a scheme for analyzing the kinematics of particles entrained by the normal fluid, to which we give the name G2, separately from those trapped on vortices, which we call G1~\cite{Mastracci2018b}. Using this separation scheme, we proposed a simple estimation of the mean free path of G2 particles through the vortex tangle, we showed that G1 velocity fluctuations are likely caused by fluctuations of the local vortex line velocity, and we showed that power-law tails in transverse particle velocity probability density functions (PDFs) are due entirely to G1. In the present paper, we expand upon these ideas, with a focus on demonstrating the utility of PTV for quantitative analysis of the vortex tangle. After a brief overview of the experimental apparatus and data analysis scheme in Sec.~\ref{sec:Protocol}, we motivate, present, and discuss each main result in its own section. An experimental estimation of the mean vortex line spacing using flow visualization is presented in Sect.~\ref{sect:resultsC}. An experimental estimation of $c_2$, a parameter related to energy dissipation in quantum turbulence~\cite{Vinen2002}, is presented in Sect.~\ref{sect:resultsB}. A direct link between vortex line reconnection and G1 transverse velocity PDF power-law tails is established in Sect.~\ref{sect:resultsA}. We conclude in Sect.~\ref{sect:conclusion}.
	
	\section{Experimental Protocol}
	\label{sec:Protocol}
\begin{figure}[bt]
        \centering
        \includegraphics[width=8.5cm,keepaspectratio]{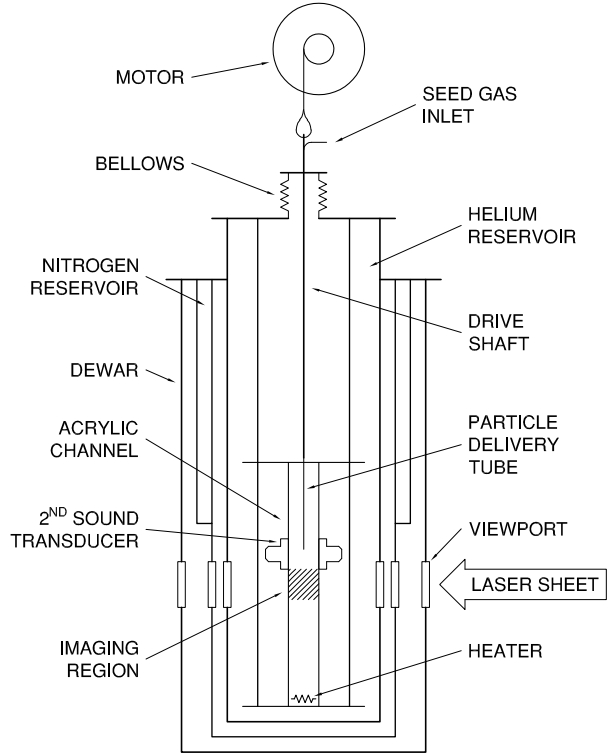}
        \caption{Simple illustration of the experimental apparatus (not to scale).\label{fig:setup}}
    \end{figure}

    This work employs the same apparatus, illustrated in Fig.~\ref{fig:setup}, described in our previous papers~\cite{Mastracci2018a,Mastracci2018b}. Solidified deuterium tracer particles with mean diameter $d_p\approx4.6~\mu$m are delivered via stainless steel tube to the center of a $1.6\times1.6\times33$~cm vertical flow channel immersed in a saturated He~II bath. The delivery tube is then retracted by an external electric motor, and a 400~$\Omega$ planar resistive heater at the bottom of the channel generates thermal counterflow. A laser beam with cross section approximately 200~$\mu$m thick and 9~mm tall illuminates particles as they move through the geometric center of the channel, and a high-speed digital camera captures them on video. A modified feature point tracking algorithm~\cite{Sbalzarini2005} yields the position of each particle in each video frame, information that can be readily transformed into velocity measurements for each particle.
    
\begin{figure}[bt]
        \centering
        \includegraphics[width=17cm,keepaspectratio]{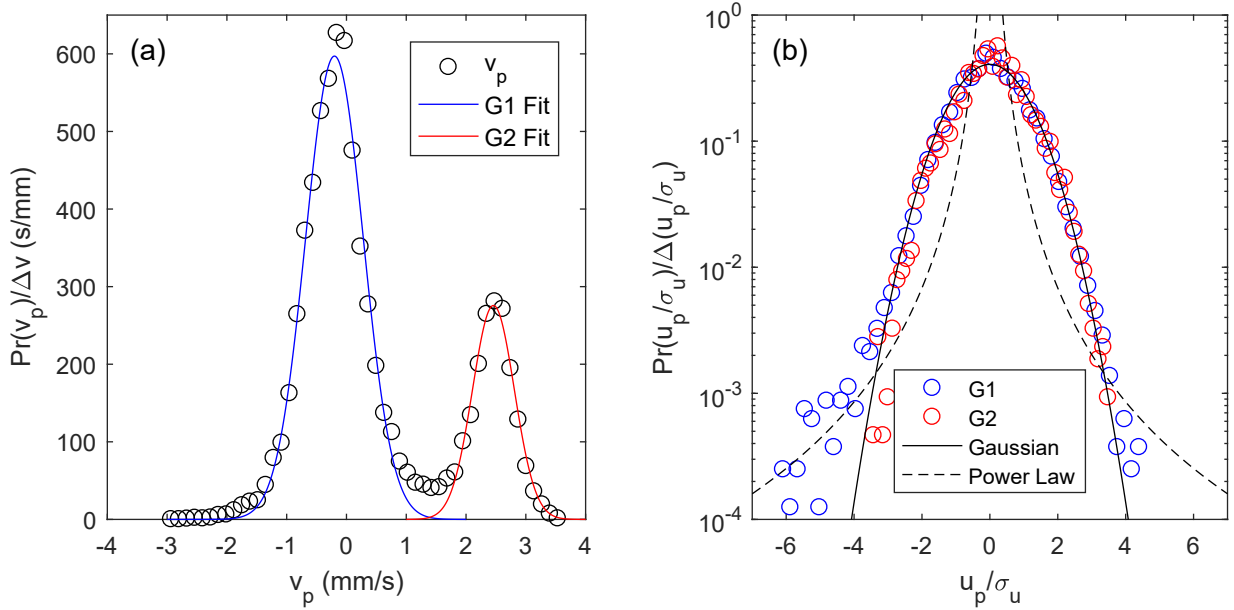}
        \caption{Typical (a) streamwise velocity PDF ($T=1.85$~K, $q=38$~mW/cm\textsuperscript{2}) and (b) transverse velocity PDFs ($T=2.00$~K, $q=113$~mW/cm\textsuperscript{2}) obtained from PTV measurements of thermal counterflow at relatively low heat flux.\label{fig:pdfs}}
    \end{figure}

    Our data set covers three temperatures, $T=1.70$, 1.85, and 2.00~K, with heat currents ranging from 29--481~mW/cm\textsuperscript{2}. Fig.~\ref{fig:pdfs} shows (a)~streamwise and (b)~transverse particle velocity PDFs typical of PTV measurements in thermal counterflow driven by relatively low heat flux. In the streamwise PDFs, one peak arises from G1, the name we give to particles trapped on quantized vortices, and the other from G2, the name we give to particles entrained by the normal fluid. To determine the category to which a velocity measurement $v_p$ contributes, we apply the following criteria~\cite{Mastracci2018b}. If $v_p<\mu_2-2\sigma_2$, where $\mu_2$ and $\sigma_2$ are the mean and standard deviation, respectively, of the G2 peak, then $v_p$ exhibits G1 behavior. If $v_p>\mu_1+2\sigma_1$, then $v_p$ exhibits G2 behavior. In the event that the peaks are well separated, i.e., $\mu_2-\mu_1>2\left(\sigma_1+\sigma_2\right)$, the criteria are reversed. The separation scheme results in ensembles of velocity measurements representing G1 and G2, which can be used for further analysis, including generation of the transverse PDFs of Fig.~\ref{fig:pdfs}(b), which are normalized by standard deviation. It can be seen that a Gaussian curve ($\mu=0$, $\sigma=1$), indicated by the solid black line, fits the core of the G1 PDF and the entirety of the G2 PDF. Beyond about four standard deviations from the center, a power law curve ($\propto\lvert{u_p}\rvert^{-3}$), indicated by the dashed black line, passes through the tail of the G1 PDF.
    
    In addition to PTV, we employ second sound attenuation to measure the average vortex line length per unit volume, or vortex line density $L$, inside the channel. As a consequence of the two fluid model, He~II supports multiple speeds of sound, including second sound, the wave-like propagation of temperature or entropy. A pair of second sound transducers, as illustrated in Fig.~\ref{fig:setup}, establish a standing second sound wave across the channel that is attenuated in the presence of quantized vortices, and the vortex line density can be obtained from the degree of attenuation~\cite{Skrbek2012}.
	
	\section{Mean free path and vortex line density}
    \label{sect:resultsC}
    
    To explain the underlying mechanism that governs whether particles exhibit G1 or G2 behavior, we acknowledged that at the beginning of each video acquisition, some particles are already trapped on vortices (G1) while others are not (G2). Untrapped particles then move over a distance comparable to their mean free path $s$ through the vortex tangle. We proposed a fairly simple formula to describe the mean free path,
    \begin{equation}
    s\leq\frac{4}{\pi{d_p}L},
    \label{eqn:mfp}
    \end{equation}
    and showed that, qualitatively, the mean free path predicted by this simple model agrees with the length of observed G2 tracks~\cite{Mastracci2018b}. To explore the usefulness of this model, we will accept its validity, and use Eqn.~(\ref{eqn:mfp}) to estimate the mean vortex line spacing $\ell=L^{-1/2}$ by using the length of G2 tracks to represent $s$.
    
    We first recognize that, for 2D planar velocimetry, G2 tracks begin and end for reasons other than de-trapping or trapping events. Particles tracing the normal fluid are free to enter and leave the imaging plane through the top or bottom of the image as well as by drifting in- or out-of-plane in the direction normal to the camera, leading to many observations of tracks that are shorter than the mean free path. It is therefore inappropriate to assume that the mean G2 track length accurately represents $s$. Alternatively, since it is not possible to observe a track longer than the mean free path (at least, not much longer), we estimate it with the mean length of the longest 10\% of observed G2 tracks. 
    
\begin{figure}[bt]
        \centering
        \includegraphics[width=9cm,keepaspectratio]{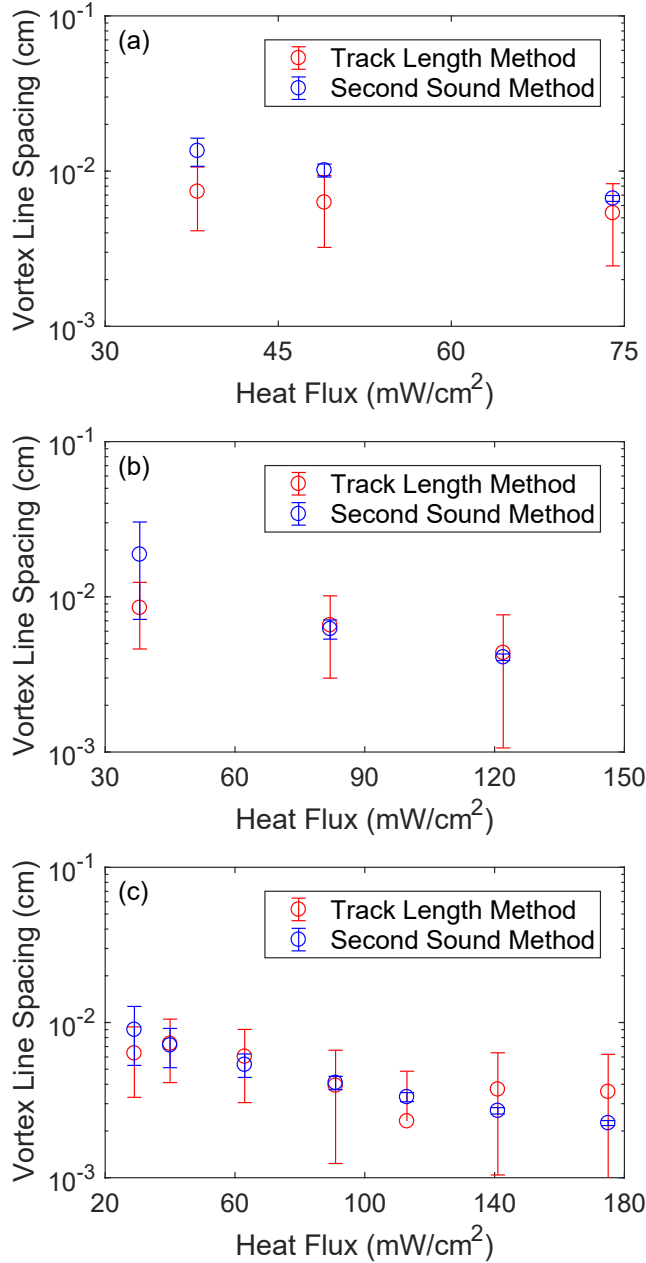}
        \caption{Prediction of mean vortex line spacing using G2 mean free path model and traditional second sound attenuation for (a)~$T=1.70$~K, (b)~$T=1.85$~K, and (c)~$T=2.00$~K.\label{fig:lineSpacing}}
    \end{figure}

    Fig.~\ref{fig:lineSpacing} shows the mean vortex line spacing as a function of heat flux for (a)~1.70~K, (b)~1.85~K, and (c)~2.00~K. Red markers predict $\ell$ using Eqn.~(\ref{eqn:mfp}), where the longest 10\% of G2 tracks observed for each point in the parameter space represent $s$ and $d_p=4.6\pm1.3$~$\mu$m~\cite{Mastracci2018a}. Blue markers represent the line spacing obtained from traditional second sound attenuation.
    
    For a simple approximation, the accuracy is remarkable, and suggests that PTV may be a viable method for estimating vortex line density in steady-state thermal counterflow. However, the assumption that G2 track lengths represent the mean free path should be approached with caution. It does not account for the possibility of a mean vortex tangle drift, an effect which is difficult to predict due to limited understanding. The true mean free path might be given as $s=s_p\left(1-C\right)$, where $s_p$ represents the observed mean free path of the particles (i.e., the mean length of the longest 10\% of observed G2 tracks), and $C=v_L/v_n$ is a correction factor relating the mean vortex tangle drift velocity $v_L$ to the normal fluid velocity. Several experiments suggest that $v_L$ is similar to $v_s$ for small enough heat flux~\cite{Wang1987,Paoletti2008b,Chagovets2011,Mastracci2018b}. We therefore constrain $C>v_s/v_n$, or equivalently, through conservation of mass, $C>-\rho_n/\rho_s$. It has also been reported that the tangle may drift in the same direction as the normal fluid when the heat flux is larger~\cite{Ashton1975}.
    
    Disparity among existing experiments makes it difficult to define a precise value for $C$, but we can infer the following picture. For counterflow driven by small heat flux, when $v_L\approx{v_s}$~\cite{Wang1987,Paoletti2008b,Chagovets2011,Mastracci2018b}, $C\approx-\rho_n/\rho_s$ and a correction factor of $\rho/\rho_s$ should be applied to Eqn.~\ref{eqn:mfp}. This may improve agreement between the curves shown in Fig.~\ref{fig:lineSpacing} for low heat flux. For higher heat flux, when $v_L<<v_n$~\cite{Awschalom1984,Wang1987}, $C\approx0$, which is consistent with the agreement shown in Fig.~\ref{fig:lineSpacing} between line spacing measured by second sound attenuation and the mean free path model for higher heat flux. 
    
    Subject to these minor corrections, the mean free path model shows strong validity as an alternative to second sound attenuation for estimation of vortex line density in steady-state thermal counterflow. Since PTV provides spatially resolved velocity measurements, this tool makes localized measurements of vortex line density possible. 
    
    \section{Experimental measurement of $c_2$}
    \label{sect:resultsB}
    
    In our recent paper, we showed that streamwise and transverse G1 velocity fluctuations $\sigma_{G1}$ as functions of counterflow velocity $v_{ns}=\lvert{v_n}\rvert+\lvert{v_s}\rvert$ can be fit remarkably well by the anticipated root mean square vortex line velocity fluctuations $\left<v_L^2\right>^{1/2}$~\cite{Mastracci2018b}. Based on local self-induced vortex motion, $\left<v_L^2\right>^{1/2}$ is given by
    \begin{equation}
    \left<v_L^2\right>^{1/2}\approx\frac{\kappa{c_2}\gamma}{4\pi}\ln\left(\frac{\ell}{\xi_0}\right)\left(\frac{\rho}{\rho_s}v_n-v_0\right)
    \label{eq:lineFluctuation}
    \end{equation}
    provided the counterflow velocity $\rho{v_n}/\rho_s$ exceeds a small critical velocity $v_0$. To illustrate agreement with $\sigma_{G1}$, we computed $\left<v_L^2\right>^{1/2}$ using values for $\gamma$ (a temperature-dependent parameter relating $L$ to $v_{ns}$), $v_0$, and the parameter $c_2$ reported in the recent work of Gao et al.~\cite{Gao2017b,Gao2018}, and we approximated the mean vortex line spacing $\ell=L^{-1/2}$ across the entire parameter space. Since the agreement between $\sigma_{G1}$ and $\left<v_L^2\right>^{1/2}$ appeared to be reasonable, we can use measured G1 velocity fluctuations to represent $\left<v_L^2\right>^{1/2}$, and apply Eqn.~(\ref{eq:lineFluctuation}) to estimate $c_2$. This parameter was introduced as a temperature-dependent coefficient relating vortex line density to local line curvature~\cite{Schwarz1988}, and has received recent attention for its role in vortex line dynamics.
    
    The parameter $c_2$ is used to describe both the build-up~\cite{Varga2018} and decay~\cite{Vinen2002} of quantum turbulence. Recent numerical simulations and experiments suggest that besides temperature, $c_2$ may depend on the specific flow geometry~\cite{Gao2018} as well as local vortex line density~\cite{Varga2018}. Since estimation of spatially-dependent $c_2$ is still beyond the capability of numerical simulations~\cite{Gao2018}, and the traditional second sound method provides averaged information across the measurement volume, application of PTV to make whole-field measurements of G1 velocity fluctuations offers a unique opportunity to investigate the spatial dependence of $c_2$. 
    
    To demonstrate estimation of $c_2$ using experimental G1 particle data, we use Eqn.~(\ref{eq:lineFluctuation}) to calculate its average value across the imaging plane. We begin by obtaining values for $\ell$, $\gamma$, and $v_0$ using our own apparatus, employing both flow visualization and second sound attenuation according to the procedures outlined by Gao et al.~\cite{Gao2017b}. The results for each temperature are tabulated in Table~\ref{table:gamma_v0}. It is unclear why $v_0<0$ at 2.00~K. 
    
    \begin{table}[h]
        \centering
        \caption{Measured values for the $\gamma$-coefficient and $v_0$.\label{table:gamma_v0}}
        \begin{tabular}{c|c|c|c}    
            T (K) &$\gamma$ (s/cm\textsuperscript{2}) &$v_0$ (cm/s) &$c_2$\\
            \hline
            1.70 &$178.6\pm42.3$ &$0.134\pm0.135$ &$0.835\pm0.239$\\
            1.85 &$236.7\pm22.5$ &$0.109\pm0.062$ &$0.563\pm0.103$\\
            2.00 &$277.6\pm11.0$ &$-0.160\pm0.038$ &$0.501\pm0.077$\\ 
        \end{tabular}
    \end{table} 
    
\begin{figure}[bt]
        \centering
        \includegraphics[width=17cm,keepaspectratio]{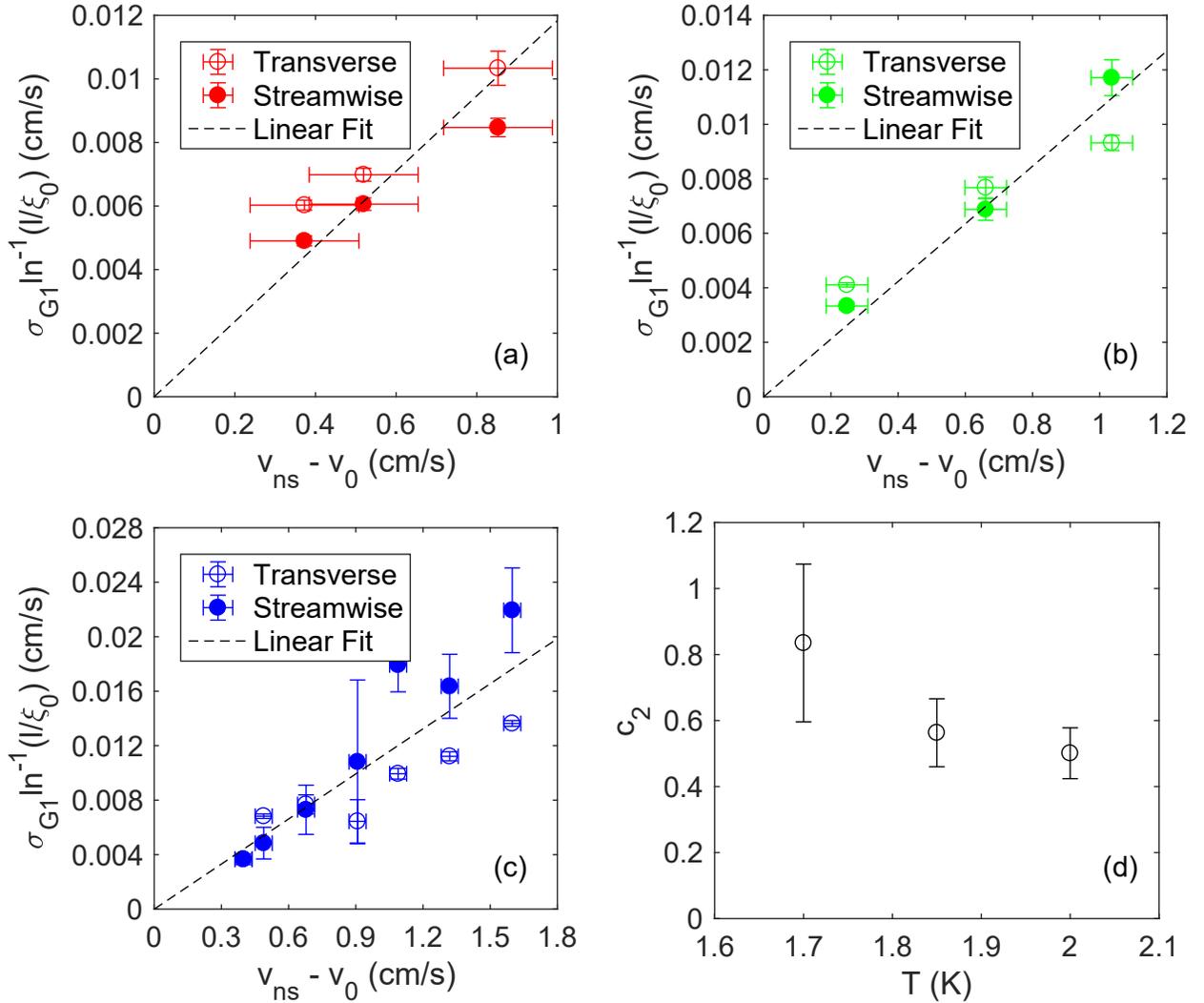}
        \caption{Linear fit to $\sigma_{G1}\ln^{-1}\left(\ell/\xi_0\right)$ as a function of $v_{ns}-v_0$ at (a) $T=1.70$, (b) 1.85, and (c) 2.00~K. (d) Extracted values of $c_2$ as a function of temperature.\label{fig:c2}}
    \end{figure}

    Values for $c_2$ can then be obtained using the procedure illustrated in Fig.~\ref{fig:c2}. Panels (a)--(c) show $\sigma_{G1}\ln^{-1}\left(\ell/\xi_0\right)$ as a function of $v_{ns}-v_0$ for $T=1.70$, 1.85, and 2.00~K, respectively. The dashed lines represent linear fits for which, according to Eqn.~(\ref{eq:lineFluctuation}), the slope is $\kappa\gamma{c_2}/4\pi$. Values for $c_2$ that produce the lines are shown in Fig.~\ref{fig:c2}(d) and tabulated in Table~\ref{table:gamma_v0}. They are slightly less than those reported in existing simulations~\cite{Schwarz1988,Gao2018} and experiments~\cite{Varga2018}, but the overall trend, a decrease with increasing temperature, is preserved. Geometric factors, i.e., the relatively large size of the experimental flow channel, may be partially responsible for the difference. It should also be kept in mind that while fluctuations of the local vortex line velocity play a large role in G1 velocity fluctuations~\cite{Mastracci2018b}, they are not solely responsible. Other factors, such as drag from the normal fluid, can also affect the G1 particle velocity~\cite{Kivotides2008c,Mineda2013}. Nonetheless, the results indicate that use of PTV to estimate $c_2$ is indeed feasible, implying that the parameter can be spatially resolved by estimating its local value based on local G1 velocity fluctuations.
    
    \section{Vortex reconnection and velocity PDF tails}
    \label{sect:resultsA}
    
    Transverse velocity $u_p$ PDFs for solidified particles tracing thermal counterflow typically exhibit a classical Gaussian core with $\left|u_p\right|^{-3}$ power law tails~\cite{Paoletti2008a,LaMantia2014a}. Recently, we applied our separation scheme to reveal that these tails can be attributed to G1~\cite{Mastracci2018b}. The tails might arise from two physical mechanisms, the superfluid velocity field or vortex reconnection, but it is not known conclusively which is responsible. 
    
    The PDF for a velocity field in the vicinity of a singular vortex is proportional to $\left|v\right|^{-3}$ for large values of the velocity, i.e., in the tail region~\cite{Min1996,Paoletti2011}. Therefore, the PDF for $v_s$ in the vicinity of a quantized vortex should exhibit the power law tails. This explanation has been invoked to explain the observation of power law tails in transverse particle velocity $u_p$ PDFs~\cite{LaMantia2014a,LaMantia2014b}. We note in passing, however, that solid particles in the vicinity of a vortex line tend to become trapped rather than respond to the superfluid itself~\cite{Barenghi2007,Kivotides2008a,Paoletti2011}. 
    
    Alternatively, when two vortices approach, reconnect, and separate from each other, the minimum separation distance $\delta$ grows in time as $\delta\propto\left|t-t_0\right|^{1/2}$, where $t_0$ is the time at which the reconnection occurs~\cite{deWaele1994,Bewley2008b,Paoletti2008a}. The separation velocity is then proportional to $\left|t-t_0\right|^{-1/2}$, and the PDF should take a form proportional to $\left|v\right|^{-3}$. Since particles have a tendency to become trapped on vortices, this scaling should be reflected in the observed motion of trapped particles and their corresponding velocity PDFs. Indeed, Paoletti et al. have shown through visualization of decaying counterflow that particle velocity PDFs take the form $\left|v\right|^{-3}$, and they identified numerous pairs of particles moving away from each other with the separation distance growing proportionally to $\left|t-t_0\right|^{1/2}$~\cite{Paoletti2008a}. This is certainly a convincing link between vortex reconnection and velocity PDF power law tails, but no direct link was established between these pairs of particles and the tail region of the PDF.
    
    With the separation scheme, a direct link can be established by analyzing the kinematics of particles that exhibit G1 behavior \textit{and} contribute to the transverse PDF tail region. Since our data comes from steady-state counterflow, acceleration along the tracks must be considered to remove effects of the mean flow. Based on the $\delta\propto\left|t-t_0\right|^{1/2}$ scaling, acceleration along tracks containing a vortrex reconnection should be proportional to $\left|t-t_0\right|^{-3/2}$. 
    
\begin{figure}[p]
        \centering
        \includegraphics[]{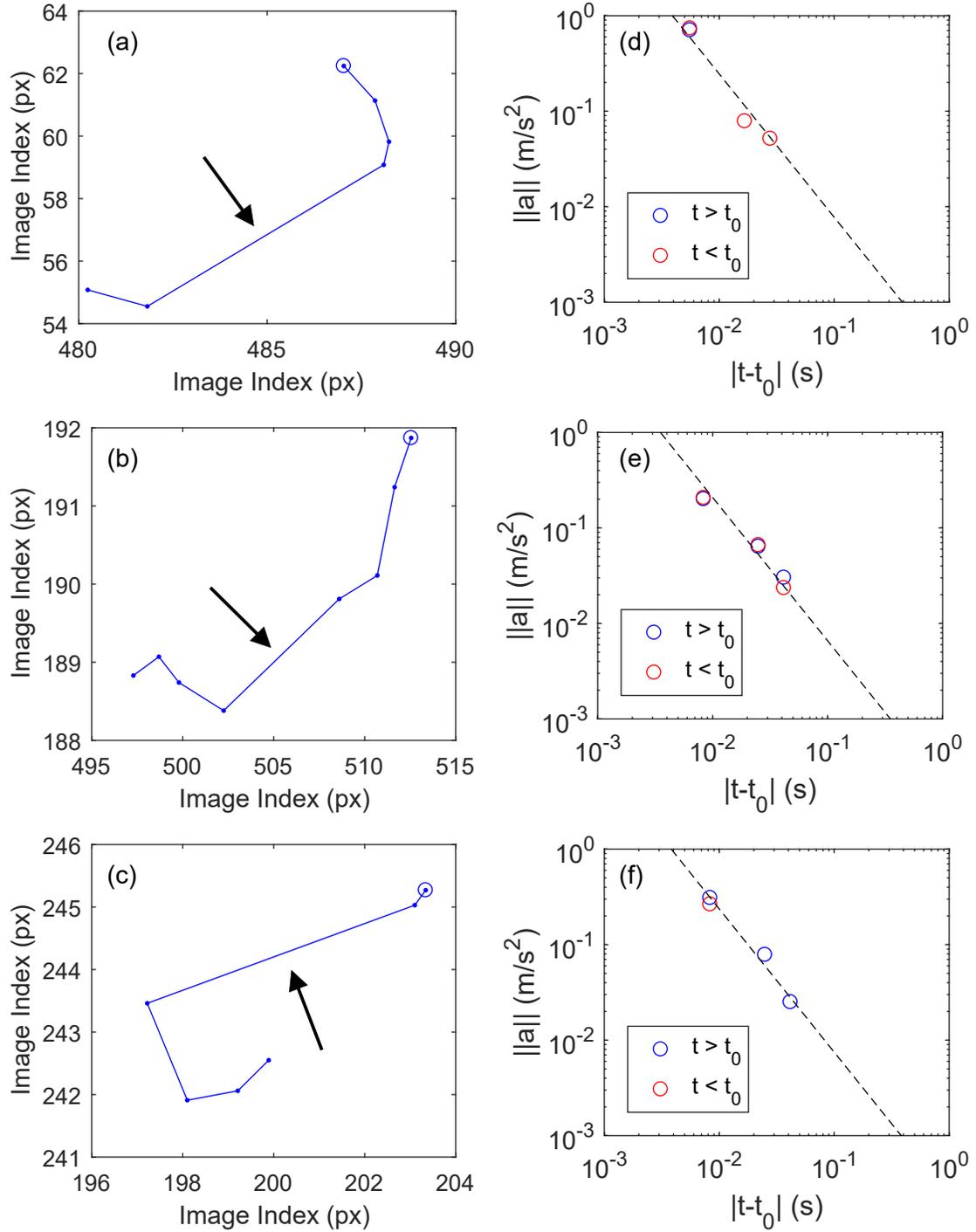}
        \caption{Selected G1 tracks that contribute to transverse PDF tails at (a)~1.70~K, (b)~1.85~K, and (c)~2.00~K. Blue circles indicate the beginning of each track and black arrows indicate the segment that contributes to the transverse velocity PDF tail. (d)--(f) Corresponding acceleration along the tracks. Dashed lines represent Eqn.~(\ref{eqn:g1Accel}).\label{fig:g1Accel}}
    \end{figure}

    We first identify G1 tracks containing a segment that contributes to the G1 transverse velocity PDF tail region, which we define as $\left|u_p\right|>\mu_{u_p}+4\sigma_{u_p}$ (see Fig.~\ref{fig:pdfs}). Figs.~\ref{fig:g1Accel}(a--c) show an example of these G1 tracks at each temperature, with the first point in each track indicated by a blue circle. In each of these tracks, the high velocity segment (indicated by the arrow) is accompanied by a strong acceleration and deceleration as well as a noticeable change in direction. These characteristics are indicative of vortex reconnection. 
    
    As a first approximation, we assume that reconnection occurs midway through the track segment that contributes to the PDF tail. It follows that the beginning of the identified segment occurs at $t=t_0-dt/2$, and the end of the segment occurs at $t=t_0+dt/2$, where $dt$ is the image acquisition interval. We can then calculate acceleration along each track away from (forward event) and towards (reverse event)~\cite{Paoletti2008a} the reconnection site as a function of elapsed time, and fit the acceleration magnitude $\lVert\mathbf{a}\rVert$ for each candidate track with a power law curve of the form
    \begin{equation}
    \left\Vert\mathbf{a}\right\Vert=C\left|t-t_0\right|^{-3/2}
    \label{eqn:g1Accel}
    \end{equation}
    where $C$ is the fitting parameter. Figs.~\ref{fig:g1Accel}(d--f) show the acceleration magnitudes along each of the corresponding tracks in Figs.~\ref{fig:g1Accel}(a--c). Forward events are shown in blue and reverse events in red, and the dashed line represents Eqn.~(\ref{eqn:g1Accel}). In all three cases, acceleration along the track agrees remarkably well with the predicted $\left|t-t_0\right|^{-3/2}$ scaling. Interestingly, the fitting parameter is approximately the same in all three cases, having an average value of $C\approx0.25$~mm/s\textsuperscript{1/2} independent of temperature. This provides a positive link between transverse velocity PDF tails and vortex reconnection, since the G1 tracks that contribute to the tails obey the acceleration scaling extrapolated from the work of Paoletti et al.~\cite{Paoletti2008a}.
    
    \section{Conclusion}
    \label{sect:conclusion}
    
    Our separation scheme for separately analyzing particles entrained by the normal fluid and those trapped on quantized vortices has led to three noteworthy observations. A simple but remarkably accurate model for the mean free path of particles traveling through the vortex tangle relates G2 track length to mean vortex line spacing, providing a new way to estimate localized vortex line density in steady-state thermal counterflow. G1 velocity fluctuations have been used to estimate the value of $c_2$, an important parameter related to dissipation of turbulent energy in He~II, using a flow visualization method that allows spatial resolution. Finally, vortex reconnection has been positively linked to particle velocity PDF power law tails by showing that acceleration along G1 tracks that contribute to the tails follows the predicted scaling for vortices accelerating away from (or towards) a reconnection site. Together, these observation indicate that with an appropriate approach to data analysis, i.e., our separation scheme, PTV is indeed a useful utility for quantifying characteristics of the vortex tangle in steady thermal counterflow.
	
	\begin{acknowledgments}
    This work was supported by U.S. Department of Energy grant DE-FG02-96ER40952. It was conducted at the National High Magnetic Field Laboratory, which was supported by NSF DMR-1157490 and DMR-1644779 and the State of Florida.
	\end{acknowledgments}
	

\begin{thebibliography}{37}%
\makeatletter
\providecommand \@ifxundefined [1]{%
 \@ifx{#1\undefined}
}%
\providecommand \@ifnum [1]{%
 \ifnum #1\expandafter \@firstoftwo
 \else \expandafter \@secondoftwo
 \fi
}%
\providecommand \@ifx [1]{%
 \ifx #1\expandafter \@firstoftwo
 \else \expandafter \@secondoftwo
 \fi
}%
\providecommand \natexlab [1]{#1}%
\providecommand \enquote  [1]{``#1''}%
\providecommand \bibnamefont  [1]{#1}%
\providecommand \bibfnamefont [1]{#1}%
\providecommand \citenamefont [1]{#1}%
\providecommand \href@noop [0]{\@secondoftwo}%
\providecommand \href [0]{\begingroup \@sanitize@url \@href}%
\providecommand \@href[1]{\@@startlink{#1}\@@href}%
\providecommand \@@href[1]{\endgroup#1\@@endlink}%
\providecommand \@sanitize@url [0]{\catcode `\\12\catcode `\$12\catcode
  `\&12\catcode `\#12\catcode `\^12\catcode `\_12\catcode `\%12\relax}%
\providecommand \@@startlink[1]{}%
\providecommand \@@endlink[0]{}%
\providecommand \url  [0]{\begingroup\@sanitize@url \@url }%
\providecommand \@url [1]{\endgroup\@href {#1}{\urlprefix }}%
\providecommand \urlprefix  [0]{URL }%
\providecommand \Eprint [0]{\href }%
\providecommand \doibase [0]{http://dx.doi.org/}%
\providecommand \selectlanguage [0]{\@gobble}%
\providecommand \bibinfo  [0]{\@secondoftwo}%
\providecommand \bibfield  [0]{\@secondoftwo}%
\providecommand \translation [1]{[#1]}%
\providecommand \BibitemOpen [0]{}%
\providecommand \bibitemStop [0]{}%
\providecommand \bibitemNoStop [0]{.\EOS\space}%
\providecommand \EOS [0]{\spacefactor3000\relax}%
\providecommand \BibitemShut  [1]{\csname bibitem#1\endcsname}%
\let\auto@bib@innerbib\@empty
\bibitem [{\citenamefont {Guo}\ \emph {et~al.}(2014)\citenamefont {Guo},
  \citenamefont {{La Mantia}}, \citenamefont {Lathrop},\ and\ \citenamefont
  {{Van Sciver}}}]{Guo2014}%
  \BibitemOpen
  \bibfield  {author} {\bibinfo {author} {\bibfnamefont {W.}~\bibnamefont
  {Guo}}, \bibinfo {author} {\bibfnamefont {M.}~\bibnamefont {{La Mantia}}},
  \bibinfo {author} {\bibfnamefont {D.~P.}\ \bibnamefont {Lathrop}}, \ and\
  \bibinfo {author} {\bibfnamefont {S.~W.}\ \bibnamefont {{Van Sciver}}},\
  }\bibfield  {title} {\enquote {\bibinfo {title} {Visualization of two-fluid
  flows of superfluid helium-4},}\ }\href {\doibase 10.1073/pnas.1312546111}
  {\bibfield  {journal} {\bibinfo  {journal} {Proc. Natl. Acad. Sci. USA}\
  }\textbf {\bibinfo {volume} {111}},\ \bibinfo {pages} {4653} (\bibinfo {year}
  {2014})}\BibitemShut {NoStop}%
\bibitem [{\citenamefont {Tisza}(1938)}]{Tisza1938}%
  \BibitemOpen
  \bibfield  {author} {\bibinfo {author} {\bibfnamefont {L.}~\bibnamefont
  {Tisza}},\ }\bibfield  {title} {\enquote {\bibinfo {title} {Transport
  phenomena in helium {II}},}\ }\href {\doibase 10.1038/141913a0} {\bibfield
  {journal} {\bibinfo  {journal} {Nature}\ }\textbf {\bibinfo {volume} {141}},\
  \bibinfo {pages} {913} (\bibinfo {year} {1938})}\BibitemShut {NoStop}%
\bibitem [{\citenamefont {Landau}(1941)}]{Landau1941}%
  \BibitemOpen
  \bibfield  {author} {\bibinfo {author} {\bibfnamefont {L.}~\bibnamefont
  {Landau}},\ }\bibfield  {title} {\enquote {\bibinfo {title} {Theory of the
  superfluidity of helium {II}},}\ }\href {\doibase 10.1103/PhysRev.60.356}
  {\bibfield  {journal} {\bibinfo  {journal} {Phys. Rev.}\ }\textbf {\bibinfo
  {volume} {60}},\ \bibinfo {pages} {356} (\bibinfo {year} {1941})}\BibitemShut
  {NoStop}%
\bibitem [{\citenamefont {Sergeev}\ and\ \citenamefont
  {Barenghi}(2009)}]{Sergeev2009}%
  \BibitemOpen
  \bibfield  {author} {\bibinfo {author} {\bibfnamefont {Y.~A.}\ \bibnamefont
  {Sergeev}}\ and\ \bibinfo {author} {\bibfnamefont {C.~F.}\ \bibnamefont
  {Barenghi}},\ }\bibfield  {title} {\enquote {\bibinfo {title}
  {Particles-vortex interactions and flow visualization in {4He}},}\ }\href
  {\doibase 10.1007/s10909-009-9994-8} {\bibfield  {journal} {\bibinfo
  {journal} {J. Low Temp. Phys.}\ }\textbf {\bibinfo {volume} {157}},\ \bibinfo
  {pages} {429} (\bibinfo {year} {2009})}\BibitemShut {NoStop}%
\bibitem [{\citenamefont {Parks}\ and\ \citenamefont
  {Donnelly}(1966)}]{Parks1966}%
  \BibitemOpen
  \bibfield  {author} {\bibinfo {author} {\bibfnamefont {P.~E.}\ \bibnamefont
  {Parks}}\ and\ \bibinfo {author} {\bibfnamefont {R.~J.}\ \bibnamefont
  {Donnelly}},\ }\bibfield  {title} {\enquote {\bibinfo {title} {Radii of
  positive and negative ions in helium {II}},}\ }\href {\doibase
  10.1103/PhysRevLett.16.45} {\bibfield  {journal} {\bibinfo  {journal} {Phys.
  Rev. Lett.}\ }\textbf {\bibinfo {volume} {16}},\ \bibinfo {pages} {45}
  (\bibinfo {year} {1966})}\BibitemShut {NoStop}%
\bibitem [{\citenamefont {Bewley}\ \emph {et~al.}(2006)\citenamefont {Bewley},
  \citenamefont {Lathrop},\ and\ \citenamefont {Sreenivasan}}]{Bewley2006}%
  \BibitemOpen
  \bibfield  {author} {\bibinfo {author} {\bibfnamefont {G.~P.}\ \bibnamefont
  {Bewley}}, \bibinfo {author} {\bibfnamefont {D.~P.}\ \bibnamefont {Lathrop}},
  \ and\ \bibinfo {author} {\bibfnamefont {K.~R.}\ \bibnamefont
  {Sreenivasan}},\ }\bibfield  {title} {\enquote {\bibinfo {title} {Superfluid
  helium: Visualization of quantized vortices},}\ }\href {\doibase
  10.1038/441588a} {\bibfield  {journal} {\bibinfo  {journal} {Nature}\
  }\textbf {\bibinfo {volume} {441}},\ \bibinfo {pages} {588} (\bibinfo {year}
  {2006})}\BibitemShut {NoStop}%
\bibitem [{\citenamefont {Kivotides}(2008)}]{Kivotides2008c}%
  \BibitemOpen
  \bibfield  {author} {\bibinfo {author} {\bibfnamefont {D.}~\bibnamefont
  {Kivotides}},\ }\bibfield  {title} {\enquote {\bibinfo {title} {Normal-fluid
  velocity measurement and superfluid vortex detection in thermal counterflow
  turbulence},}\ }\href {\doibase 10.1103/PhysRevB.78.224501} {\bibfield
  {journal} {\bibinfo  {journal} {Phys. Rev. B}\ }\textbf {\bibinfo {volume}
  {78}},\ \bibinfo {pages} {224501} (\bibinfo {year} {2008})}\BibitemShut
  {NoStop}%
\bibitem [{\citenamefont {Mineda}\ \emph {et~al.}(2013)\citenamefont {Mineda},
  \citenamefont {Tsubota}, \citenamefont {Sergeev}, \citenamefont {Barenghi},\
  and\ \citenamefont {Vinen}}]{Mineda2013}%
  \BibitemOpen
  \bibfield  {author} {\bibinfo {author} {\bibfnamefont {Y.}~\bibnamefont
  {Mineda}}, \bibinfo {author} {\bibfnamefont {M.}~\bibnamefont {Tsubota}},
  \bibinfo {author} {\bibfnamefont {Y.~A.}\ \bibnamefont {Sergeev}}, \bibinfo
  {author} {\bibfnamefont {C.~F.}\ \bibnamefont {Barenghi}}, \ and\ \bibinfo
  {author} {\bibfnamefont {W.~F.}\ \bibnamefont {Vinen}},\ }\bibfield  {title}
  {\enquote {\bibinfo {title} {Velocity distributions of tracer particles in
  thermal counterflow in superfluid ${}^{4}${He}},}\ }\href {\doibase
  10.1103/PhysRevB.87.174508} {\bibfield  {journal} {\bibinfo  {journal} {Phys.
  Rev. B}\ }\textbf {\bibinfo {volume} {87}},\ \bibinfo {pages} {174508}
  (\bibinfo {year} {2013})}\BibitemShut {NoStop}%
\bibitem [{\citenamefont {Marakov}\ \emph {et~al.}(2015)\citenamefont
  {Marakov}, \citenamefont {Gao}, \citenamefont {Guo}, \citenamefont
  {Van~Sciver}, \citenamefont {Ihas}, \citenamefont {McKinsey},\ and\
  \citenamefont {Vinen}}]{Marakov2015}%
  \BibitemOpen
  \bibfield  {author} {\bibinfo {author} {\bibfnamefont {A.}~\bibnamefont
  {Marakov}}, \bibinfo {author} {\bibfnamefont {J.}~\bibnamefont {Gao}},
  \bibinfo {author} {\bibfnamefont {W.}~\bibnamefont {Guo}}, \bibinfo {author}
  {\bibfnamefont {S.~W.}\ \bibnamefont {Van~Sciver}}, \bibinfo {author}
  {\bibfnamefont {G.~G.}\ \bibnamefont {Ihas}}, \bibinfo {author}
  {\bibfnamefont {D.~N.}\ \bibnamefont {McKinsey}}, \ and\ \bibinfo {author}
  {\bibfnamefont {W.~F.}\ \bibnamefont {Vinen}},\ }\bibfield  {title} {\enquote
  {\bibinfo {title} {Visualization of the normal-fluid turbulence in
  counterflowing superfluid $^{4}\mathrm{He}$},}\ }\href {\doibase
  10.1103/PhysRevB.91.094503} {\bibfield  {journal} {\bibinfo  {journal} {Phys.
  Rev. B}\ }\textbf {\bibinfo {volume} {91}},\ \bibinfo {pages} {094503}
  (\bibinfo {year} {2015})}\BibitemShut {NoStop}%
\bibitem [{\citenamefont {Vinen}(1957)}]{VinenIII}%
  \BibitemOpen
  \bibfield  {author} {\bibinfo {author} {\bibfnamefont {W.~F.}\ \bibnamefont
  {Vinen}},\ }\bibfield  {title} {\enquote {\bibinfo {title} {Mutual friction
  in a heat current in liquid helium {II}. {III}. {T}heory of the mutual
  friction},}\ }\href {\doibase 10.1098/rspa.1957.0191} {\bibfield  {journal}
  {\bibinfo  {journal} {Proc. R. Soc. London, Ser. A}\ }\textbf {\bibinfo
  {volume} {242}},\ \bibinfo {pages} {493} (\bibinfo {year}
  {1957})}\BibitemShut {NoStop}%
\bibitem [{\citenamefont {Hall}\ and\ \citenamefont {Vinen}(1956)}]{Hall1956b}%
  \BibitemOpen
  \bibfield  {author} {\bibinfo {author} {\bibfnamefont {H.~E.}\ \bibnamefont
  {Hall}}\ and\ \bibinfo {author} {\bibfnamefont {W.~F.}\ \bibnamefont
  {Vinen}},\ }\bibfield  {title} {\enquote {\bibinfo {title} {The rotation of
  liquid helium {II II. The} theory of mutual friction in uniformly rotating
  helium {II}},}\ }\href {\doibase 10.1098/rspa.1956.0215} {\bibfield
  {journal} {\bibinfo  {journal} {Proc. R. Soc. London Ser. A}\ }\textbf
  {\bibinfo {volume} {238}},\ \bibinfo {pages} {215} (\bibinfo {year}
  {1956})}\BibitemShut {NoStop}%
\bibitem [{\citenamefont {Guo}\ \emph {et~al.}(2013)\citenamefont {Guo},
  \citenamefont {McKinsey}, \citenamefont {Marakov}, \citenamefont {Thompson},
  \citenamefont {Ihas},\ and\ \citenamefont {Vinen}}]{Guo2013}%
  \BibitemOpen
  \bibfield  {author} {\bibinfo {author} {\bibfnamefont {W.}~\bibnamefont
  {Guo}}, \bibinfo {author} {\bibfnamefont {D.~N.}\ \bibnamefont {McKinsey}},
  \bibinfo {author} {\bibfnamefont {A.}~\bibnamefont {Marakov}}, \bibinfo
  {author} {\bibfnamefont {K.~J.}\ \bibnamefont {Thompson}}, \bibinfo {author}
  {\bibfnamefont {G.~G.}\ \bibnamefont {Ihas}}, \ and\ \bibinfo {author}
  {\bibfnamefont {W.~F.}\ \bibnamefont {Vinen}},\ }\bibfield  {title} {\enquote
  {\bibinfo {title} {Visualization technique for determining the structure
  functions of normal-fluid turbulence in superfluid helium-4},}\ }\href
  {\doibase 10.1007/s10909-012-0708-2} {\bibfield  {journal} {\bibinfo
  {journal} {J. Low Temp. Phys.}\ }\textbf {\bibinfo {volume} {171}},\ \bibinfo
  {pages} {497} (\bibinfo {year} {2013})}\BibitemShut {NoStop}%
\bibitem [{\citenamefont {Vinen}(2014)}]{Vinen2014}%
  \BibitemOpen
  \bibfield  {author} {\bibinfo {author} {\bibfnamefont {W.~F.}\ \bibnamefont
  {Vinen}},\ }\bibfield  {title} {\enquote {\bibinfo {title} {Quantum
  turbulence: Aspects of visualization and homogeneous turbulence},}\ }\href
  {\doibase 10.1007/s10909-013-0911-9} {\bibfield  {journal} {\bibinfo
  {journal} {J. Low Temp. Phys.}\ }\textbf {\bibinfo {volume} {175}},\ \bibinfo
  {pages} {305} (\bibinfo {year} {2014})}\BibitemShut {NoStop}%
\bibitem [{\citenamefont {Gao}\ \emph {et~al.}(2015)\citenamefont {Gao},
  \citenamefont {Marakov}, \citenamefont {Guo}, \citenamefont {Pawlowski},
  \citenamefont {{Van Sciver}}, \citenamefont {Ihas}, \citenamefont
  {McKinsey},\ and\ \citenamefont {Vinen}}]{Gao2015}%
  \BibitemOpen
  \bibfield  {author} {\bibinfo {author} {\bibfnamefont {J.}~\bibnamefont
  {Gao}}, \bibinfo {author} {\bibfnamefont {A.}~\bibnamefont {Marakov}},
  \bibinfo {author} {\bibfnamefont {W.}~\bibnamefont {Guo}}, \bibinfo {author}
  {\bibfnamefont {B.~T.}\ \bibnamefont {Pawlowski}}, \bibinfo {author}
  {\bibfnamefont {S.~W.}\ \bibnamefont {{Van Sciver}}}, \bibinfo {author}
  {\bibfnamefont {G.~G.}\ \bibnamefont {Ihas}}, \bibinfo {author}
  {\bibfnamefont {D.~N.}\ \bibnamefont {McKinsey}}, \ and\ \bibinfo {author}
  {\bibfnamefont {W.~F.}\ \bibnamefont {Vinen}},\ }\bibfield  {title} {\enquote
  {\bibinfo {title} {Producing and imaging a thin line of {He2∗} molecular
  tracers in helium-4},}\ }\href {\doibase 10.1063/1.4930147} {\bibfield
  {journal} {\bibinfo  {journal} {Rev. Sci. Instrum.}\ }\textbf {\bibinfo
  {volume} {86}},\ \bibinfo {pages} {093904} (\bibinfo {year}
  {2015})}\BibitemShut {NoStop}%
\bibitem [{\citenamefont {Chagovets}\ and\ \citenamefont {{Van
  Sciver}}(2011)}]{Chagovets2011}%
  \BibitemOpen
  \bibfield  {author} {\bibinfo {author} {\bibfnamefont {T.~V.}\ \bibnamefont
  {Chagovets}}\ and\ \bibinfo {author} {\bibfnamefont {S.~W.}\ \bibnamefont
  {{Van Sciver}}},\ }\bibfield  {title} {\enquote {\bibinfo {title} {A study of
  thermal counterflow using particle tracking velocimetry},}\ }\href {\doibase
  10.1063/1.3657084} {\bibfield  {journal} {\bibinfo  {journal} {Phys. Fluids}\
  }\textbf {\bibinfo {volume} {23}},\ \bibinfo {pages} {107102} (\bibinfo
  {year} {2011})}\BibitemShut {NoStop}%
\bibitem [{\citenamefont {{La Mantia}}\ and\ \citenamefont
  {Skrbek}(2014)}]{LaMantia2014a}%
  \BibitemOpen
  \bibfield  {author} {\bibinfo {author} {\bibfnamefont {M.}~\bibnamefont {{La
  Mantia}}}\ and\ \bibinfo {author} {\bibfnamefont {L.}~\bibnamefont
  {Skrbek}},\ }\bibfield  {title} {\enquote {\bibinfo {title} {Quantum, or
  classical turbulence?}}\ }\href {\doibase 10.1209/0295-5075/105/46002}
  {\bibfield  {journal} {\bibinfo  {journal} {Europhys. Lett.}\ }\textbf
  {\bibinfo {volume} {105}},\ \bibinfo {pages} {46002} (\bibinfo {year}
  {2014})}\BibitemShut {NoStop}%
\bibitem [{\citenamefont {La~Mantia}\ and\ \citenamefont
  {Skrbek}(2014)}]{LaMantia2014b}%
  \BibitemOpen
  \bibfield  {author} {\bibinfo {author} {\bibfnamefont {M.}~\bibnamefont
  {La~Mantia}}\ and\ \bibinfo {author} {\bibfnamefont {L.}~\bibnamefont
  {Skrbek}},\ }\bibfield  {title} {\enquote {\bibinfo {title} {Quantum
  turbulence visualized by particle dynamics},}\ }\href {\doibase
  10.1103/PhysRevB.90.014519} {\bibfield  {journal} {\bibinfo  {journal} {Phys.
  Rev. B}\ }\textbf {\bibinfo {volume} {90}},\ \bibinfo {pages} {014519}
  (\bibinfo {year} {2014})}\BibitemShut {NoStop}%
\bibitem [{\citenamefont {Mastracci}\ and\ \citenamefont
  {Guo}(2018{\natexlab{a}})}]{Mastracci2018b}%
  \BibitemOpen
  \bibfield  {author} {\bibinfo {author} {\bibfnamefont {B.}~\bibnamefont
  {Mastracci}}\ and\ \bibinfo {author} {\bibfnamefont {W.}~\bibnamefont
  {Guo}},\ }\bibfield  {title} {\enquote {\bibinfo {title} {Exploration of
  thermal counterflow in {He II} using particle tracking velocimetry},}\ }\href
  {\doibase 10.1103/PhysRevFluids.3.063304} {\bibfield  {journal} {\bibinfo
  {journal} {Phys. Rev. Fluids}\ }\textbf {\bibinfo {volume} {3}},\ \bibinfo
  {pages} {063304} (\bibinfo {year} {2018}{\natexlab{a}})}\BibitemShut
  {NoStop}%
\bibitem [{\citenamefont {Vinen}\ and\ \citenamefont
  {Niemela}(2002)}]{Vinen2002}%
  \BibitemOpen
  \bibfield  {author} {\bibinfo {author} {\bibfnamefont {W.~F.}\ \bibnamefont
  {Vinen}}\ and\ \bibinfo {author} {\bibfnamefont {J.~J.}\ \bibnamefont
  {Niemela}},\ }\bibfield  {title} {\enquote {\bibinfo {title} {Quantum
  turbulence},}\ }\href {\doibase 10.1023/A:1019695418590} {\bibfield
  {journal} {\bibinfo  {journal} {J. Low Temp. Phys.}\ }\textbf {\bibinfo
  {volume} {128}},\ \bibinfo {pages} {167} (\bibinfo {year}
  {2002})}\BibitemShut {NoStop}%
\bibitem [{\citenamefont {Mastracci}\ and\ \citenamefont
  {Guo}(2018{\natexlab{b}})}]{Mastracci2018a}%
  \BibitemOpen
  \bibfield  {author} {\bibinfo {author} {\bibfnamefont {B.}~\bibnamefont
  {Mastracci}}\ and\ \bibinfo {author} {\bibfnamefont {W.}~\bibnamefont
  {Guo}},\ }\bibfield  {title} {\enquote {\bibinfo {title} {An apparatus for
  generation and quantitative measurement of homogeneous isotropic turbulence
  in {He II}},}\ }\href {\doibase 10.1063/1.4997735} {\bibfield  {journal}
  {\bibinfo  {journal} {Rev. Sci. Instrum.}\ }\textbf {\bibinfo {volume}
  {89}},\ \bibinfo {pages} {015107} (\bibinfo {year}
  {2018}{\natexlab{b}})}\BibitemShut {NoStop}%
\bibitem [{\citenamefont {Sbalzarini}\ and\ \citenamefont
  {Koumoutsakos}(2005)}]{Sbalzarini2005}%
  \BibitemOpen
  \bibfield  {author} {\bibinfo {author} {\bibfnamefont {I.~F.}\ \bibnamefont
  {Sbalzarini}}\ and\ \bibinfo {author} {\bibfnamefont {P.}~\bibnamefont
  {Koumoutsakos}},\ }\bibfield  {title} {\enquote {\bibinfo {title} {Feature
  point tracking and trajectory analysis for video imaging in cell biology},}\
  }\href {\doibase https://doi.org/10.1016/j.jsb.2005.06.002} {\bibfield
  {journal} {\bibinfo  {journal} {J. Struct. Biol.}\ }\textbf {\bibinfo
  {volume} {151}},\ \bibinfo {pages} {182} (\bibinfo {year}
  {2005})}\BibitemShut {NoStop}%
\bibitem [{\citenamefont {Skrbek}\ and\ \citenamefont
  {Sreenivasan}(2012)}]{Skrbek2012}%
  \BibitemOpen
  \bibfield  {author} {\bibinfo {author} {\bibfnamefont {L.}~\bibnamefont
  {Skrbek}}\ and\ \bibinfo {author} {\bibfnamefont {K.~R.}\ \bibnamefont
  {Sreenivasan}},\ }\bibfield  {title} {\enquote {\bibinfo {title} {Developed
  quantum turbulence and its decay},}\ }\href {\doibase 10.1063/1.3678335}
  {\bibfield  {journal} {\bibinfo  {journal} {Phys. Fluids}\ }\textbf {\bibinfo
  {volume} {24}},\ \bibinfo {pages} {011301} (\bibinfo {year}
  {2012})}\BibitemShut {NoStop}%
\bibitem [{\citenamefont {Wang}\ \emph {et~al.}(1987)\citenamefont {Wang},
  \citenamefont {Swanson},\ and\ \citenamefont {Donnelly}}]{Wang1987}%
  \BibitemOpen
  \bibfield  {author} {\bibinfo {author} {\bibfnamefont {R.~T.}\ \bibnamefont
  {Wang}}, \bibinfo {author} {\bibfnamefont {C.~E.}\ \bibnamefont {Swanson}}, \
  and\ \bibinfo {author} {\bibfnamefont {R.~J.}\ \bibnamefont {Donnelly}},\
  }\bibfield  {title} {\enquote {\bibinfo {title} {Anisotropy and drift of a
  vortex tangle in helium {II}},}\ }\href {\doibase 10.1103/PhysRevB.36.5240}
  {\bibfield  {journal} {\bibinfo  {journal} {Phys. Rev. B}\ }\textbf {\bibinfo
  {volume} {36}},\ \bibinfo {pages} {5240} (\bibinfo {year}
  {1987})}\BibitemShut {NoStop}%
\bibitem [{\citenamefont {Paoletti}\ \emph
  {et~al.}(2008{\natexlab{a}})\citenamefont {Paoletti}, \citenamefont
  {Fiorito}, \citenamefont {Sreenivasan},\ and\ \citenamefont
  {Lathrop}}]{Paoletti2008b}%
  \BibitemOpen
  \bibfield  {author} {\bibinfo {author} {\bibfnamefont {M.~S.}\ \bibnamefont
  {Paoletti}}, \bibinfo {author} {\bibfnamefont {R.~B.}\ \bibnamefont
  {Fiorito}}, \bibinfo {author} {\bibfnamefont {K.~R.}\ \bibnamefont
  {Sreenivasan}}, \ and\ \bibinfo {author} {\bibfnamefont {D.~P.}\ \bibnamefont
  {Lathrop}},\ }\bibfield  {title} {\enquote {\bibinfo {title} {Visualization
  of superfluid helium flow},}\ }\href {\doibase 10.1143/JPSJ.77.111007}
  {\bibfield  {journal} {\bibinfo  {journal} {J. Phys. Soc. Jpn.}\ }\textbf
  {\bibinfo {volume} {77}},\ \bibinfo {pages} {111007} (\bibinfo {year}
  {2008}{\natexlab{a}})}\BibitemShut {NoStop}%
\bibitem [{\citenamefont {Ashton}\ and\ \citenamefont
  {Northby}(1975)}]{Ashton1975}%
  \BibitemOpen
  \bibfield  {author} {\bibinfo {author} {\bibfnamefont {R.~A.}\ \bibnamefont
  {Ashton}}\ and\ \bibinfo {author} {\bibfnamefont {J.~A.}\ \bibnamefont
  {Northby}},\ }\bibfield  {title} {\enquote {\bibinfo {title} {Vortex velocity
  in turbulent {He II} counterflow},}\ }\href {\doibase
  10.1103/PhysRevLett.35.1714} {\bibfield  {journal} {\bibinfo  {journal}
  {Phys. Rev. Lett.}\ }\textbf {\bibinfo {volume} {35}},\ \bibinfo {pages}
  {1714} (\bibinfo {year} {1975})}\BibitemShut {NoStop}%
\bibitem [{\citenamefont {Awschalom}\ \emph {et~al.}(1984)\citenamefont
  {Awschalom}, \citenamefont {Milliken},\ and\ \citenamefont
  {Schwarz}}]{Awschalom1984}%
  \BibitemOpen
  \bibfield  {author} {\bibinfo {author} {\bibfnamefont {D.~D.}\ \bibnamefont
  {Awschalom}}, \bibinfo {author} {\bibfnamefont {F.~P.}\ \bibnamefont
  {Milliken}}, \ and\ \bibinfo {author} {\bibfnamefont {K.~W.}\ \bibnamefont
  {Schwarz}},\ }\bibfield  {title} {\enquote {\bibinfo {title} {Properties of
  superfluid turbulence in a large channel},}\ }\href {\doibase
  10.1103/PhysRevLett.53.1372} {\bibfield  {journal} {\bibinfo  {journal}
  {Phys. Rev. Lett.}\ }\textbf {\bibinfo {volume} {53}},\ \bibinfo {pages}
  {1372} (\bibinfo {year} {1984})}\BibitemShut {NoStop}%
\bibitem [{\citenamefont {Gao}\ \emph {et~al.}(2017)\citenamefont {Gao},
  \citenamefont {Varga}, \citenamefont {Guo},\ and\ \citenamefont
  {Vinen}}]{Gao2017b}%
  \BibitemOpen
  \bibfield  {author} {\bibinfo {author} {\bibfnamefont {J.}~\bibnamefont
  {Gao}}, \bibinfo {author} {\bibfnamefont {E.}~\bibnamefont {Varga}}, \bibinfo
  {author} {\bibfnamefont {W.}~\bibnamefont {Guo}}, \ and\ \bibinfo {author}
  {\bibfnamefont {W.~F.}\ \bibnamefont {Vinen}},\ }\bibfield  {title} {\enquote
  {\bibinfo {title} {Energy spectrum of thermal counterflow turbulence in
  superfluid helium-4},}\ }\href {\doibase 10.1103/PhysRevB.96.094511}
  {\bibfield  {journal} {\bibinfo  {journal} {Phys. Rev. B}\ }\textbf {\bibinfo
  {volume} {96}},\ \bibinfo {pages} {094511} (\bibinfo {year}
  {2017})}\BibitemShut {NoStop}%
\bibitem [{\citenamefont {Gao}\ \emph {et~al.}(2018)\citenamefont {Gao},
  \citenamefont {Guo}, \citenamefont {Yui}, \citenamefont {Tsubota},\ and\
  \citenamefont {Vinen}}]{Gao2018}%
  \BibitemOpen
  \bibfield  {author} {\bibinfo {author} {\bibfnamefont {J.}~\bibnamefont
  {Gao}}, \bibinfo {author} {\bibfnamefont {W.}~\bibnamefont {Guo}}, \bibinfo
  {author} {\bibfnamefont {S.}~\bibnamefont {Yui}}, \bibinfo {author}
  {\bibfnamefont {M.}~\bibnamefont {Tsubota}}, \ and\ \bibinfo {author}
  {\bibfnamefont {W.~F.}\ \bibnamefont {Vinen}},\ }\bibfield  {title} {\enquote
  {\bibinfo {title} {Dissipation in quantum turbulence in superfluid
  $^{4}\mathrm{He}$ above 1 k},}\ }\href {\doibase 10.1103/PhysRevB.97.184518}
  {\bibfield  {journal} {\bibinfo  {journal} {Phys. Rev. B}\ }\textbf {\bibinfo
  {volume} {97}},\ \bibinfo {pages} {184518} (\bibinfo {year}
  {2018})}\BibitemShut {NoStop}%
\bibitem [{\citenamefont {Schwarz}(1988)}]{Schwarz1988}%
  \BibitemOpen
  \bibfield  {author} {\bibinfo {author} {\bibfnamefont {K.~W.}\ \bibnamefont
  {Schwarz}},\ }\bibfield  {title} {\enquote {\bibinfo {title}
  {Three-dimensional vortex dynamics in superfluid $^{4}\mathrm{He}$:
  Homogeneous superfluid turbulence},}\ }\href {\doibase
  10.1103/PhysRevB.38.2398} {\bibfield  {journal} {\bibinfo  {journal} {Phys.
  Rev. B}\ }\textbf {\bibinfo {volume} {38}},\ \bibinfo {pages} {2398}
  (\bibinfo {year} {1988})}\BibitemShut {NoStop}%
\bibitem [{\citenamefont {Varga}\ and\ \citenamefont
  {Skrbek}(2018)}]{Varga2018}%
  \BibitemOpen
  \bibfield  {author} {\bibinfo {author} {\bibfnamefont {E.}~\bibnamefont
  {Varga}}\ and\ \bibinfo {author} {\bibfnamefont {L.}~\bibnamefont {Skrbek}},\
  }\bibfield  {title} {\enquote {\bibinfo {title} {Dynamics of the density of
  quantized vortex lines in counterflow turbulence: {Experimental}
  investigation},}\ }\href {\doibase 10.1103/PhysRevB.97.064507} {\bibfield
  {journal} {\bibinfo  {journal} {Phys. Rev. B}\ }\textbf {\bibinfo {volume}
  {97}},\ \bibinfo {pages} {064507} (\bibinfo {year} {2018})}\BibitemShut
  {NoStop}%
\bibitem [{\citenamefont {Paoletti}\ \emph
  {et~al.}(2008{\natexlab{b}})\citenamefont {Paoletti}, \citenamefont {Fisher},
  \citenamefont {Sreenivasan},\ and\ \citenamefont {Lathrop}}]{Paoletti2008a}%
  \BibitemOpen
  \bibfield  {author} {\bibinfo {author} {\bibfnamefont {M.~S.}\ \bibnamefont
  {Paoletti}}, \bibinfo {author} {\bibfnamefont {M.~E.}\ \bibnamefont
  {Fisher}}, \bibinfo {author} {\bibfnamefont {K.~R.}\ \bibnamefont
  {Sreenivasan}}, \ and\ \bibinfo {author} {\bibfnamefont {D.~P.}\ \bibnamefont
  {Lathrop}},\ }\bibfield  {title} {\enquote {\bibinfo {title} {Velocity
  statistics distinguish quantum turbulence from classical turbulence},}\
  }\href {\doibase 10.1103/PhysRevLett.101.154501} {\bibfield  {journal}
  {\bibinfo  {journal} {Phys. Rev. Lett.}\ }\textbf {\bibinfo {volume} {101}},\
  \bibinfo {pages} {154501} (\bibinfo {year} {2008}{\natexlab{b}})}\BibitemShut
  {NoStop}%
\bibitem [{\citenamefont {Min}\ \emph {et~al.}(1996)\citenamefont {Min},
  \citenamefont {Mezić},\ and\ \citenamefont {Leonard}}]{Min1996}%
  \BibitemOpen
  \bibfield  {author} {\bibinfo {author} {\bibfnamefont {I.~A.}\ \bibnamefont
  {Min}}, \bibinfo {author} {\bibfnamefont {I.}~\bibnamefont {Mezić}}, \ and\
  \bibinfo {author} {\bibfnamefont {A.}~\bibnamefont {Leonard}},\ }\bibfield
  {title} {\enquote {\bibinfo {title} {L\'{e}vy stable distributions for
  velocity and velocity difference in systems of vortex elements},}\ }\href
  {\doibase 10.1063/1.868908} {\bibfield  {journal} {\bibinfo  {journal} {Phys.
  Fluids}\ }\textbf {\bibinfo {volume} {8}},\ \bibinfo {pages} {1169} (\bibinfo
  {year} {1996})}\BibitemShut {NoStop}%
\bibitem [{\citenamefont {Paoletti}\ and\ \citenamefont
  {Lathrop}(2011)}]{Paoletti2011}%
  \BibitemOpen
  \bibfield  {author} {\bibinfo {author} {\bibfnamefont {M.~S.}\ \bibnamefont
  {Paoletti}}\ and\ \bibinfo {author} {\bibfnamefont {D.~P.}\ \bibnamefont
  {Lathrop}},\ }\bibfield  {title} {\enquote {\bibinfo {title} {Quantum
  turbulence},}\ }\href {\doibase 10.1146/annurev-conmatphys-062910-140533}
  {\bibfield  {journal} {\bibinfo  {journal} {Annu. Rev. Con. Mat. Phys.}\
  }\textbf {\bibinfo {volume} {2}},\ \bibinfo {pages} {213} (\bibinfo {year}
  {2011})}\BibitemShut {NoStop}%
\bibitem [{\citenamefont {Barenghi}\ \emph {et~al.}(2007)\citenamefont
  {Barenghi}, \citenamefont {Kivotides},\ and\ \citenamefont
  {Sergeev}}]{Barenghi2007}%
  \BibitemOpen
  \bibfield  {author} {\bibinfo {author} {\bibfnamefont {C.~F.}\ \bibnamefont
  {Barenghi}}, \bibinfo {author} {\bibfnamefont {D.}~\bibnamefont {Kivotides}},
  \ and\ \bibinfo {author} {\bibfnamefont {Y.~A.}\ \bibnamefont {Sergeev}},\
  }\bibfield  {title} {\enquote {\bibinfo {title} {Close approach of a
  spherical particle and a quantised vortex in helium {II}},}\ }\href {\doibase
  10.1007/s10909-007-9387-9} {\bibfield  {journal} {\bibinfo  {journal} {J. Low
  Temp. Phys.}\ }\textbf {\bibinfo {volume} {148}},\ \bibinfo {pages} {293}
  (\bibinfo {year} {2007})}\BibitemShut {NoStop}%
\bibitem [{\citenamefont {Kivotides}\ \emph {et~al.}(2008)\citenamefont
  {Kivotides}, \citenamefont {Barenghi},\ and\ \citenamefont
  {Sergeev}}]{Kivotides2008a}%
  \BibitemOpen
  \bibfield  {author} {\bibinfo {author} {\bibfnamefont {D.}~\bibnamefont
  {Kivotides}}, \bibinfo {author} {\bibfnamefont {C.~F.}\ \bibnamefont
  {Barenghi}}, \ and\ \bibinfo {author} {\bibfnamefont {Y.~A.}\ \bibnamefont
  {Sergeev}},\ }\bibfield  {title} {\enquote {\bibinfo {title} {Interactions
  between particles and quantized vortices in superfluid helium},}\ }\href
  {\doibase 10.1103/PhysRevB.77.014527} {\bibfield  {journal} {\bibinfo
  {journal} {Phys. Rev. B}\ }\textbf {\bibinfo {volume} {77}},\ \bibinfo
  {pages} {014527} (\bibinfo {year} {2008})}\BibitemShut {NoStop}%
\bibitem [{\citenamefont {de~Waele}\ and\ \citenamefont
  {Aarts}(1994)}]{deWaele1994}%
  \BibitemOpen
  \bibfield  {author} {\bibinfo {author} {\bibfnamefont {A.~T. A.~M.}\
  \bibnamefont {de~Waele}}\ and\ \bibinfo {author} {\bibfnamefont {R.~G.
  K.~M.}\ \bibnamefont {Aarts}},\ }\bibfield  {title} {\enquote {\bibinfo
  {title} {Route to vortex reconnection},}\ }\href {\doibase
  10.1103/PhysRevLett.72.482} {\bibfield  {journal} {\bibinfo  {journal} {Phys.
  Rev. Lett.}\ }\textbf {\bibinfo {volume} {72}},\ \bibinfo {pages} {482}
  (\bibinfo {year} {1994})}\BibitemShut {NoStop}%
\bibitem [{\citenamefont {Bewley}\ \emph {et~al.}(2008)\citenamefont {Bewley},
  \citenamefont {Paoletti}, \citenamefont {Sreenivasan},\ and\ \citenamefont
  {Lathrop}}]{Bewley2008b}%
  \BibitemOpen
  \bibfield  {author} {\bibinfo {author} {\bibfnamefont {G.~P.}\ \bibnamefont
  {Bewley}}, \bibinfo {author} {\bibfnamefont {M.~S.}\ \bibnamefont
  {Paoletti}}, \bibinfo {author} {\bibfnamefont {K.~R.}\ \bibnamefont
  {Sreenivasan}}, \ and\ \bibinfo {author} {\bibfnamefont {D.~P.}\ \bibnamefont
  {Lathrop}},\ }\bibfield  {title} {\enquote {\bibinfo {title}
  {Characterization of reconnecting vortices in superfluid helium},}\ }\href
  {\doibase 10.1073/pnas.0806002105} {\bibfield  {journal} {\bibinfo  {journal}
  {Proc. Natl. Acad. Sci. USA}\ }\textbf {\bibinfo {volume} {105}},\ \bibinfo
  {pages} {13707} (\bibinfo {year} {2008})}\BibitemShut {NoStop}%
\end{thebibliography}
%

\end{document}